# A Simple Sonic Mapping Method Verified by CT Scan Images

**Jimmy Xuekai Li[1*], Thomas Flottmann[2], Max Millen[2], Shuai Chen[1], Yixiao Huang[1], Zhongwei Chen[1*]**

[1]The University of Queensland, St Lucia, QLD 4072, Australia

[2]Origin Energy, 180 Ann Street, Brisbane, QLD 4000, Australia

Corresponding author: Jimmy X. Li (Jimmy.Li@uq.edu.au); Zhongwei Chen (zhongwei.chen@uq.edu.au)

**Key Points:**

- A novel sonic mapping technique is used to measure both P-wave and S-wave velocities across coal samples, providing a non-destructive approach for assessing internal mechanical properties and structural heterogeneity.

- The study develops 2D sonic maps for each orthogonal direction (X, Y, Z) and refines these through interpolation, leading to the creation of a comprehensive 3D sonic map that captures the internal structure of coal samples.

- The accuracy of the 3D sonic mapping is validated by high-resolution CT scan images, confirming the reliability of the method for detecting fractures, voids, and material composition variations within the coal.

**Plain Language Summary**

- This method uses sound waves to look inside materials like coal without cutting them open or causing damage. It could be applied to other types of rocks, building materials, or even industrial components to check their internal structure.

- By measuring sound waves in different directions, we can create detailed 3D maps that show what's happening inside a material. This technique could be useful in fields like construction, geology, and material science, helping to spot weaknesses or hidden cracks.

- The accuracy of these sound-based maps is confirmed with CT scans, which are commonly used in medical imaging. This shows that the method is reliable and could be adapted to many other areas that need non-destructive testing, such as aerospace, manufacturing, or archaeology.

**Abstract**

This study presents a novel sonic mapping method applied to coal samples, verified by CT scan imaging. Cubic coal samples with side lengths of 50-70 mm were subjected to non-destructive sonic tests, measuring both P-wave (Vp) and S-wave (Vs) velocities. Each of the three orthogonal directions (X, Y, and Z) of the cube was divided into 9 regions, resulting in 27 Vp and 27 Vs measurements per sample. From these data, 2D sonic maps were constructed for each direction, and interpolation was employed to refine the mappings. A 3D sonic map was then generated by averaging the 2D maps. The 3D sonic mapping results were compared and validated against high-resolution CT scan images, confirming the reliability of this approach for mapping the internal structure of coal samples.

**1 Introduction**

Understanding the internal structure and mechanical properties of coal is crucial for various applications, including coal seam gas production (Ahamed et al., 2019; L. Liu et al., 2024; Sampath et al., 2019), mining (D. Liu et al., 2022; Yu et al., 2018), and geotechnical engineering (Kang et al., 2023; Song & Zhang, 2021). Non-destructive testing methods have gained significant attention for their ability to provide insights into material properties without causing damage (Gupta et al., 2022; Kumar & Mahto, 2013; Wang et al., 2020). Among these methods, sonic testing is particularly effective, as it allows for the measurement of both compressional (P-wave) and shear (S-wave) velocities, which are directly related to material stiffness, porosity, and anisotropy. However, to further enhance the precision of such measurements, it is essential to develop advanced mapping techniques that capture the heterogeneity of the material at a fine scale.

In this study, we propose a simple yet effective sonic mapping method to characterize the internal structure of coal. The method involves dividing a cubic coal sample into a grid of measurement points and obtaining sonic wave velocities (Vp and Vs) in three orthogonal directions (X, Y, and Z). By using this approach, we aim to construct both 2D and 3D sonic velocity maps, which provide a spatial representation of the sample's internal properties. The 2D mapping offers valuable information for each plane, while the 3D mapping, generated by averaging the 2D maps, provides a comprehensive understanding of the coal's internal structure.

To verify the accuracy of the sonic mapping results, we employed computed tomography (CT) scanning, a well-established imaging technique that provides high-resolution 3D visualizations of material density variations. CT scanning has been widely used in rock mechanics and material science for non-destructive imaging, making it an ideal method for validating our sonic mapping approach.

The combination of sonic testing and CT imaging offers a powerful framework for assessing coal's internal properties. Sonic velocities (Vp and Vs) are sensitive to the material's mechanical properties, while CT scanning provides a direct measure of its density distribution. By comparing the 3D sonic map with the CT scan images, we can evaluate the reliability of our sonic mapping method and its potential to detect heterogeneities such as fractures, voids, and variations in material composition within coal samples.

This study provides a detailed description of the experimental procedures, data processing, and analysis methods used to generate and refine both 2D and 3D sonic maps. Additionally, the effectiveness of the proposed method is demonstrated through a comprehensive comparison between the sonic maps and CT scan images. The results highlight the potential of the sonic mapping technique as a valuable tool for characterizing coal and other geological materials.

## 2 Methods

Coal samples were selected and cut into cubic shapes with side lengths ranging from 50 mm to 70 mm. The samples were carefully prepared to ensure smooth, defect-free surfaces that would not interfere with sonic wave measurements. Each cube was marked along its three orthogonal axes (X, Y, and Z) to define measurement areas. A non-destructive sonic testing method was then employed to measure both compressional (P-wave, Vp) and shear (S-wave, Vs) velocities. Each side of the cube was divided into 9 equal areas, providing a grid of 27 measurement points per wave type. Sonic transducers were placed at each grid point to transmit and receive both P-wave and S-wave signals in the X, Y, and Z directions. This arrangement resulted in 27 P-wave (Vp) and 27 S-wave (Vs) velocity values for each sample, effectively covering the cube's internal structure in three dimensions (***Figure 1 a, b***). The velocity data were used to produce comprehensive 2D maps along each principal direction, which were then averaged to form a 3D sonic map of each coal sample (***Figure 1 c***). The average value of the sub-volumetric representative is calculated by **Eq. 1**. This 3D map visualized the distribution of sonic velocities within the coal, revealing potential areas of heterogeneity or material variations.

$$E_d(avg) = \frac{E_x+E_y+E_z}{3} \qquad (1)$$

where,

$E_d(avg)$: average dynamic modulus value of the sub-volumetric representative

$E_x$: X-direction dynamic modulus value of the sub-volumetric representative

$E_y$: Y-direction dynamic modulus value of the sub-volumetric representative

$E_z$: Z-direction dynamic modulus value of the sub-volumetric representative

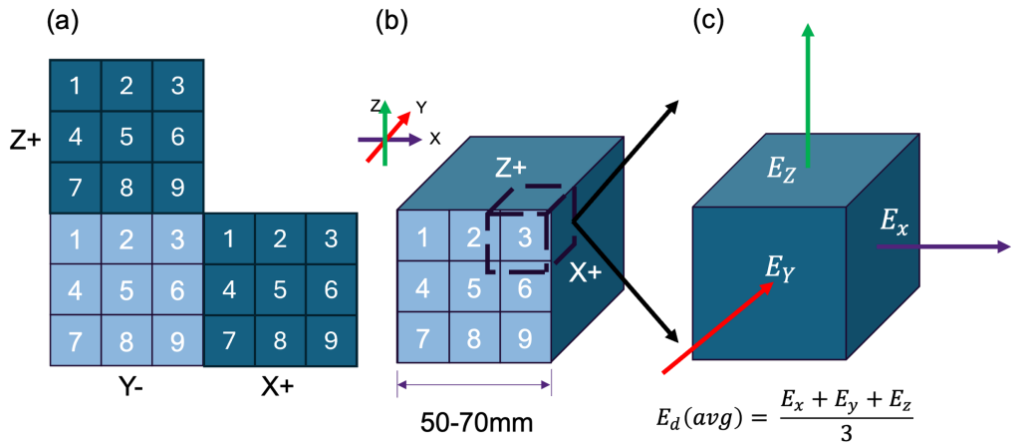

***Figure 1*** *(a) Illustration showing the division of each side of the cube into nine equal areas. (b) Representation of each side corresponding to the X, Y, and Z directions of the cube sample. (c)*

*Calculation of the dynamic modulus ($E_d$) for the sub-volume, derived as the average of the values from all three directional measurements.*

The measurement points are also shown on a 3-D digital reconstruction of a coal sample, obtained from micro-CT scans, to clearly depict the specific locations used for velocity analysis within the sample as illustrated in *Figure 2*.

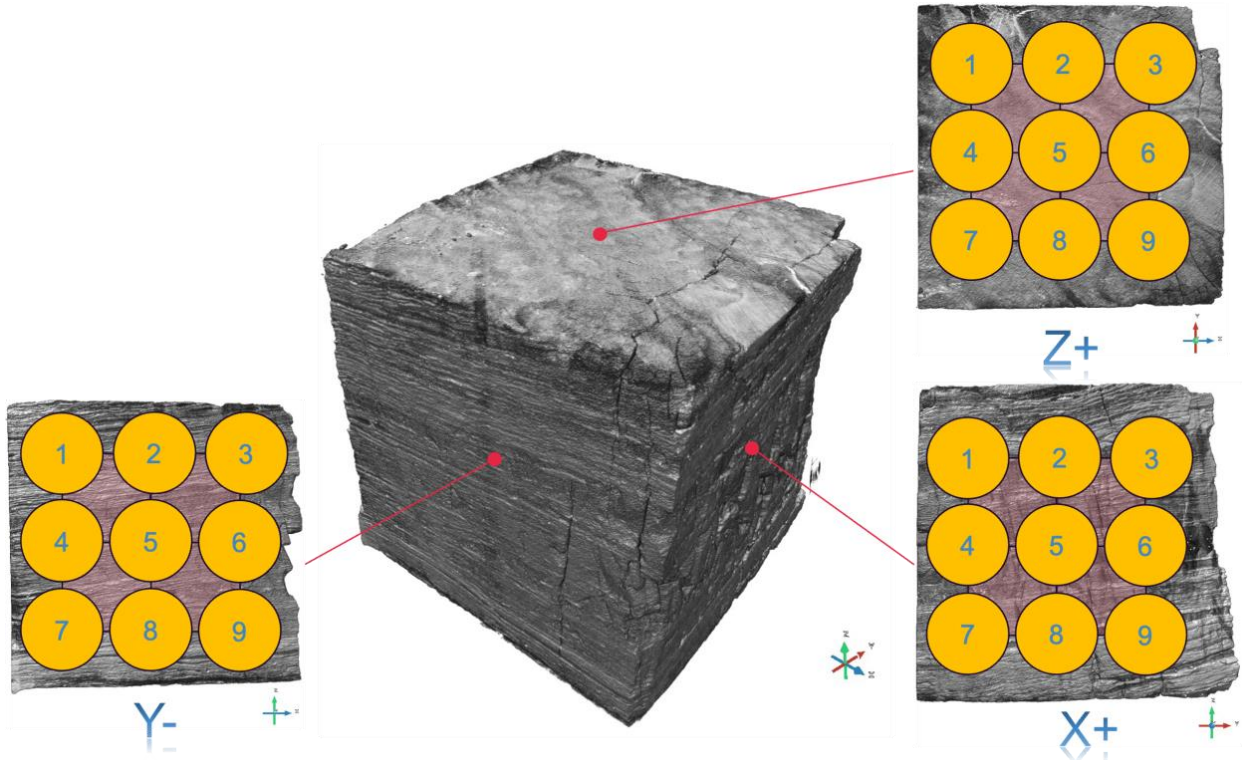

*Figure 2 Each face of the cube was segmented into nine equal parts for velocity measurement, resulting in a grid layout with 27 measurement points. These points are displayed on a 3-D reconstructed digital coal sample derived from micro-CT scans, illustrating the precise locations for velocity analysis within the sample.*

For each sample, velocity data from the X, Y, and Z directions were used to create 2D ultrasonic mappings for each plane of the cube. The interpolation algorithms were then applied to these mappings to enhance resolution, providing smoother transitions between measurement points and an improved representation of the coal's internal structure.

To further validate the ultrasonic mapping, computed tomography (CT) scans of the coal samples were conducted, generating high-resolution 3D images of the internal density distribution. This CT data served as a reference for assessing the accuracy of the sonic mapping in identifying internal features.

## 3 Results

The detailed mappings derived from the velocity measurements are illustrated in **Figure 3**. The mapped P-wave velocities (Vp) and S-wave velocities (Vs) across the cube's three orthogonal directions, showing the variation of velocities across different faces of the sample, which aids in understanding the sample's mechanical behavior under wave propagation. The dynamic Young's modulus (Ed) is also mapped, calculated from the measured Vp and Vs by using **Eq. 2**, highlighting the mechanical stiffness variations across the cube sample (Li et al., 2024). Average values for each direction are also calculated and displayed in the titles of the respective sub-diagrams.

$$E_d = \frac{\rho V_s^2 (3V_p^2 - 4V_s^2)}{V_p^2 - V_s^2} \quad (2)$$

$E_d$: dynamic Young's modulus

$\rho$: density of the sample

$V_p$: P-wave velocity

$V_s$: S-wave velocity

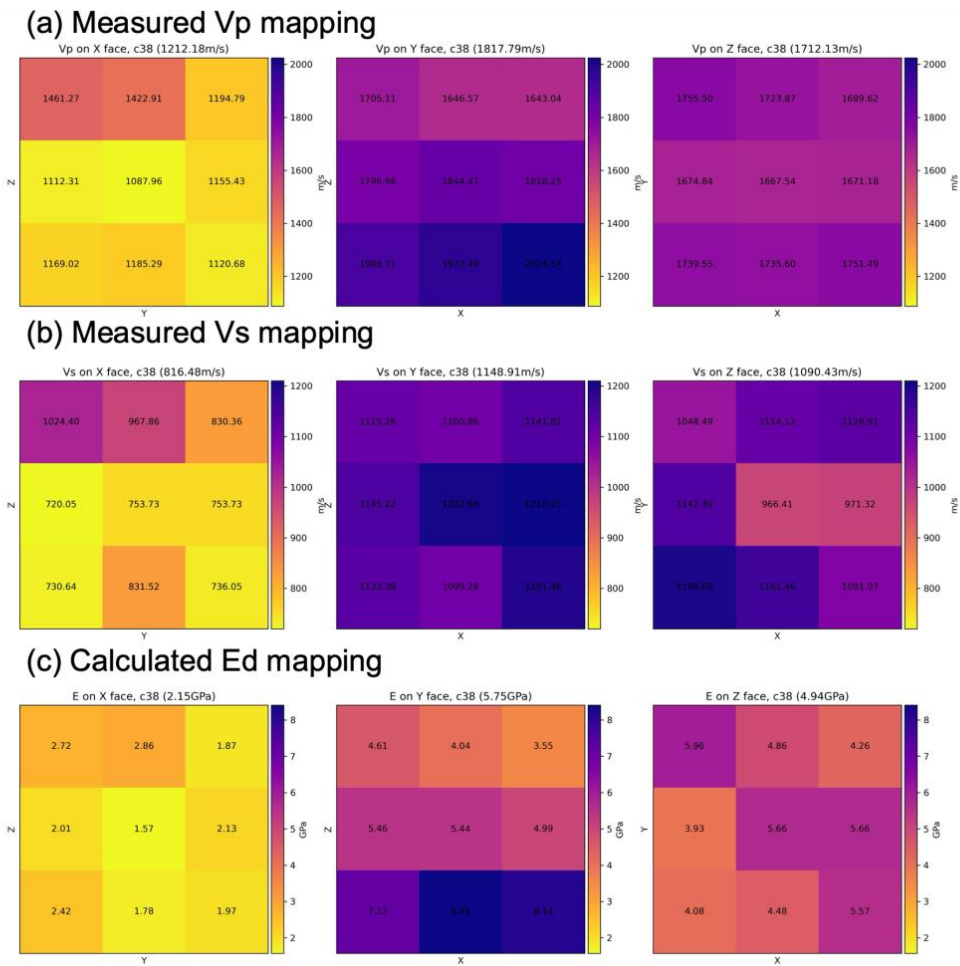

***Figure 3*** *(a) Mapping of measured P-wave velocity ($V_p$) across the X, Y, and Z directions (faces) of the cube. (b) Mapping of measured S-wave velocity ($V_s$) across the same directions. (c) Mapping of calculated dynamic Young's modulus ($E_d$) on the X, Y, and Z directions. Average values for each direction are calculated and displayed in the titles of the corresponding sub-diagrams.*

Areas displaying lower sonic velocities typically coincide with fractures or voids within the coal matrix, as indicated in the mappings. Conversely, regions exhibiting higher sonic velocities correspond to denser or more structurally intact sections of the sample. These observations are crucial for understanding the spatial distribution of material properties and their implications for the sample's mechanical behavior under stress.

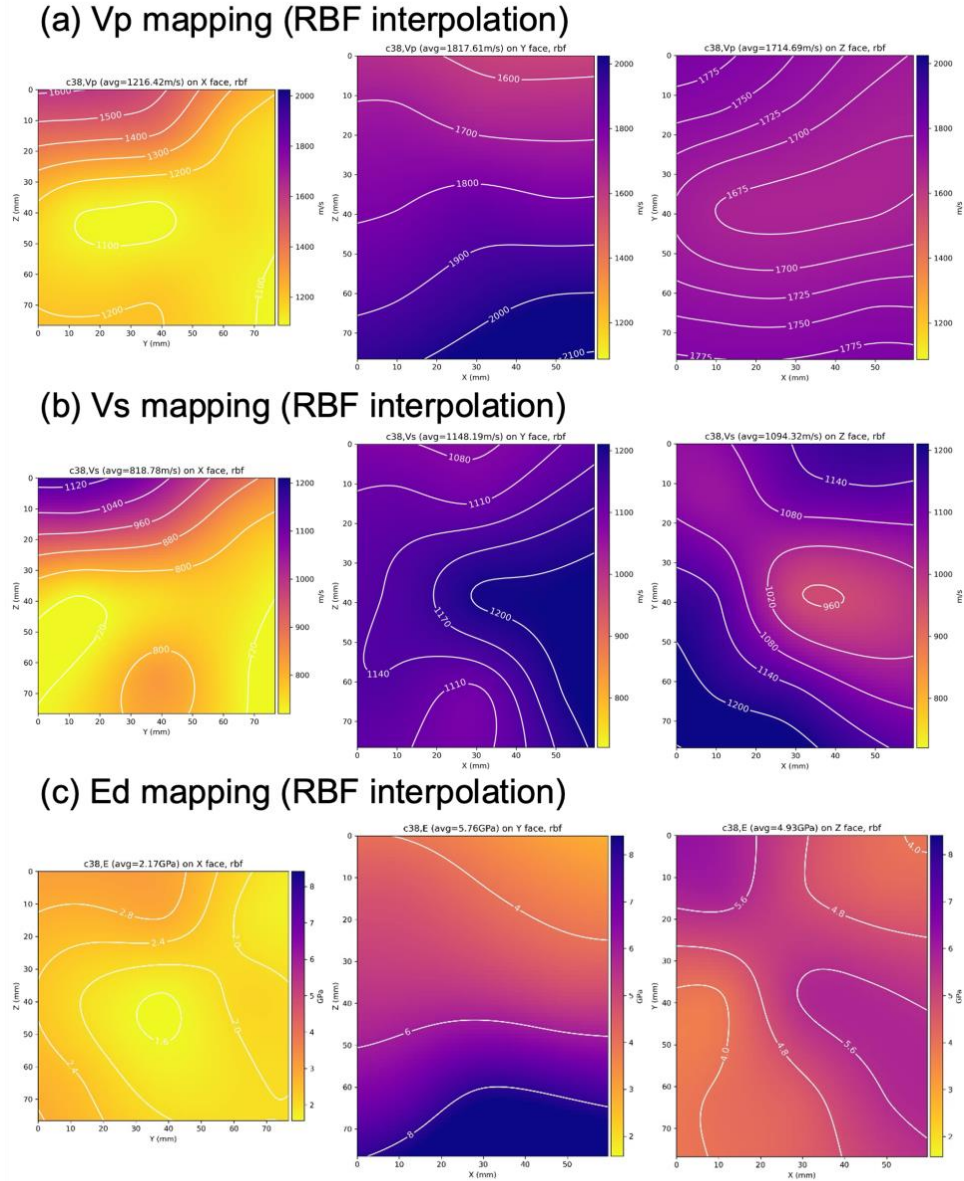

***Figure 4*** *(a) Interpolated P-wave velocity ($V_P$) contour mapping using Radial Basis Function (RBF) interpolation. (b) Interpolated S-wave velocity ($V_s$) contour mapping, also using RBF interpolation. (c) Dynamic Young's modulus ($E_d$) contour mapping on the X, Y, and Z directions (faces) of the cube, calculated by using the interpolated Vp and Vs values.*

Various interpolation methods, including radial basis function (RBF), linear regression, cubic, and Kriging interpolations, were utilized to process P-wave and S-wave velocity measurements. These techniques are selected for their ability to provide accurate and detailed velocity fields from discrete data points, each offering unique strengths in handling spatial data.

For the purpose of detailed analysis in this study, RBF interpolation was specifically chosen for its robustness in creating smooth, continuous functions, which is demonstrated in the contour mappings. **Figure 4** provides a detailed visualization of the interpolated data using the RBF method.

The mappings are crucial for understanding the isotropic or anisotropic nature of the material properties within the coal sample. By analyzing the velocities and calculated modulus, insights were gained into how the material responds to stress and deformation. This analysis highlights potential zones of weakness or strength within the sample and underscores the importance of directional dependencies in mechanical properties, which are critical for both theoretical analysis and practical application in material science and geotechnical engineering.

**Figure 5** features contour mappings of the dynamic modulus (Ed) superimposed over the 3-D digital coal sample, providing a visual juxtaposition that highlights the relationship between mechanical properties and structural integrity. This approach allows for a comparative analysis between the 3D sonic mappings and CT scan images, facilitating an in-depth assessment of the correlations between sonic velocity variations and structural features identified by the CT scans.

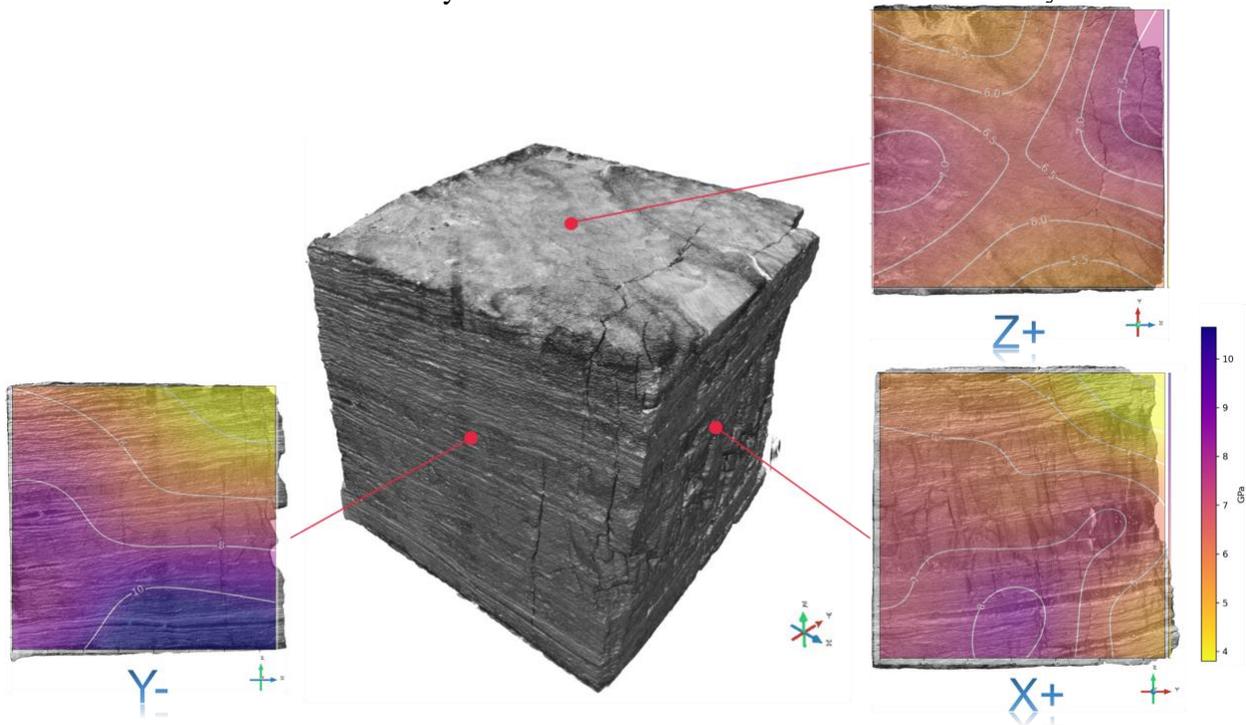

*Figure 5 Contour mappings of the dynamic modulus (Ed) are juxtapositioned over the 3-D digital coal sample.*

The extensive analysis conducted on the volumetric dynamic modulus data of the coal sample as illustrated in **Figure 6**. The 3D mapping of this data across the entire sample, providing a comprehensive view of the modulus distribution within the bulk material. **Figure 6 b** complements this by presenting a 3D contour mapping of the same data, which effectively highlights the variations and gradations in dynamic modulus values throughout the sample. These 3D mappings provide a simple new method to reveal the heterogeneous nature of the coal's mechanical properties, with clear delineations where modulus values change, indicating potential variations in material composition.

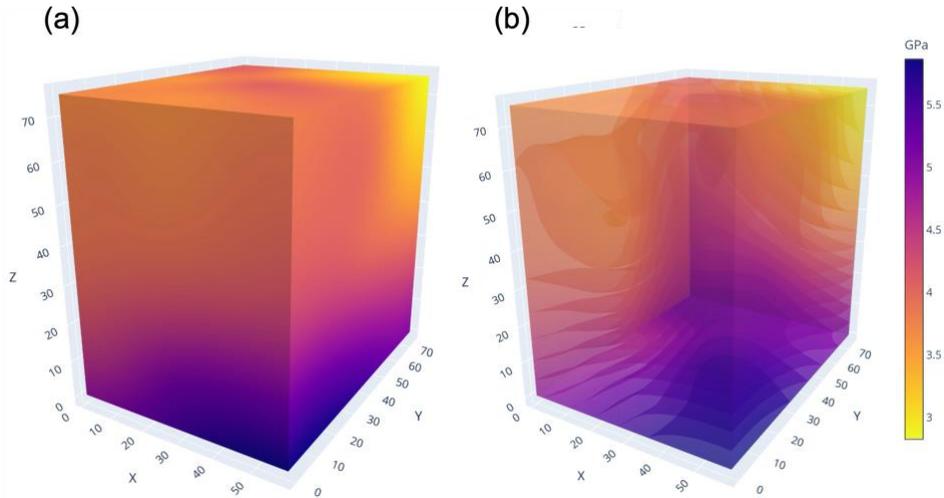

***Figure 6*** *(a) 3D mapping of the volumetric dynamic modulus data across the whole sample. (b) 3D contour mapping of the same data, highlighting the variations and distribution of dynamic modulus values throughout the sample.*

**Figure 7** provides a detailed comparison of the reconstructed digital coal sample with ultrasonic mapping, shown in both 3D and 2D views for each direction (Example 1). This visualization effectively contrasts the different imaging modalities and their ability to depict internal structures within the sample. Notably, in this example 1, the dynamic modulus mapping identifies a high-density mineral presence at the bottom of the sample, which corresponds to a lighter color in the CT scan images. This correlation underscores the effectiveness of integrating multiple imaging techniques to assess variations in material density and structural integrity across the sample.

Another example is demonstrated in Figure 8 that depicts a side-by-side comparison of the reconstructed digital coal sample with dynamic modulus mapping in both 3D and 2D representations for each direction (Example 2). This comparison uniquely demonstrates how the mapping captures the presence of a high-density mineral located at the top area of the sample, which correlates with lighter colors observed in the CT scan images. Similar to previous observations, areas with lower Ed are typically associated with fractures or voids in the coal matrix, while regions with higher sonic velocities indicate denser minerals or more structurally intact materials. This example further underscores the value of using multiple imaging modalities to reveal the spatial distribution of material properties. These details are critical for assessing the mechanical behavior of the sample under various stress conditions, offering insights into how internal structural variations can influence the overall integrity and performance of the material.

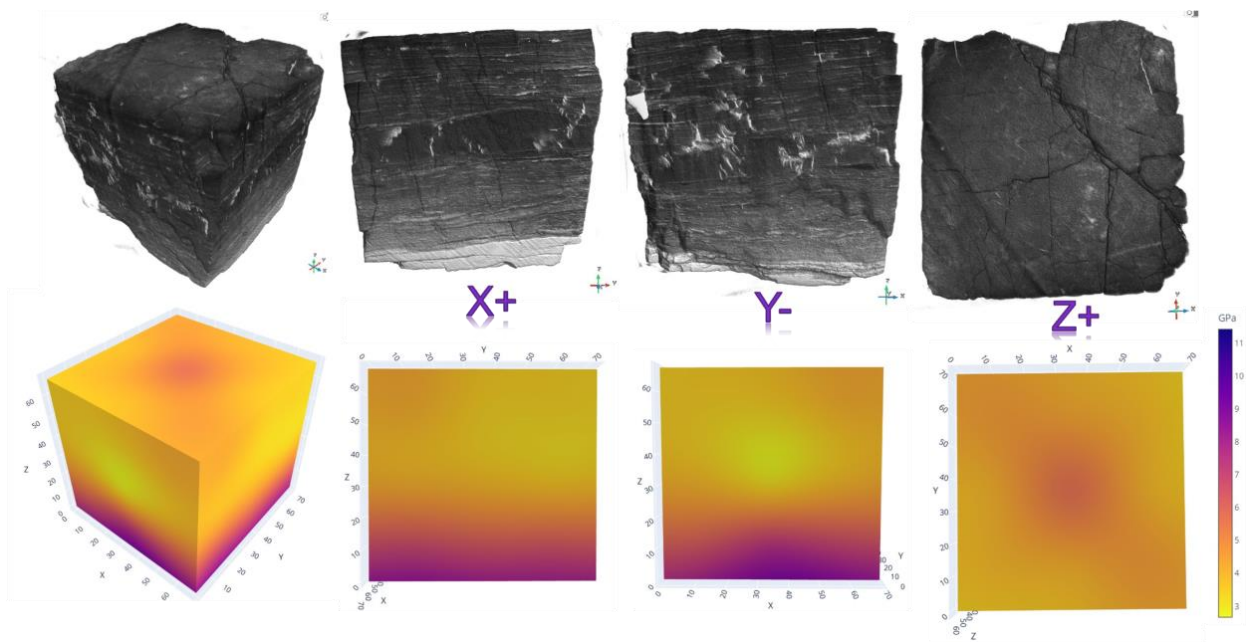

*Figure 7* *Comparison of the reconstructed digital coal sample and dynamic modulus mapping in both 3D and 2D representations for each direction (Example 1).*

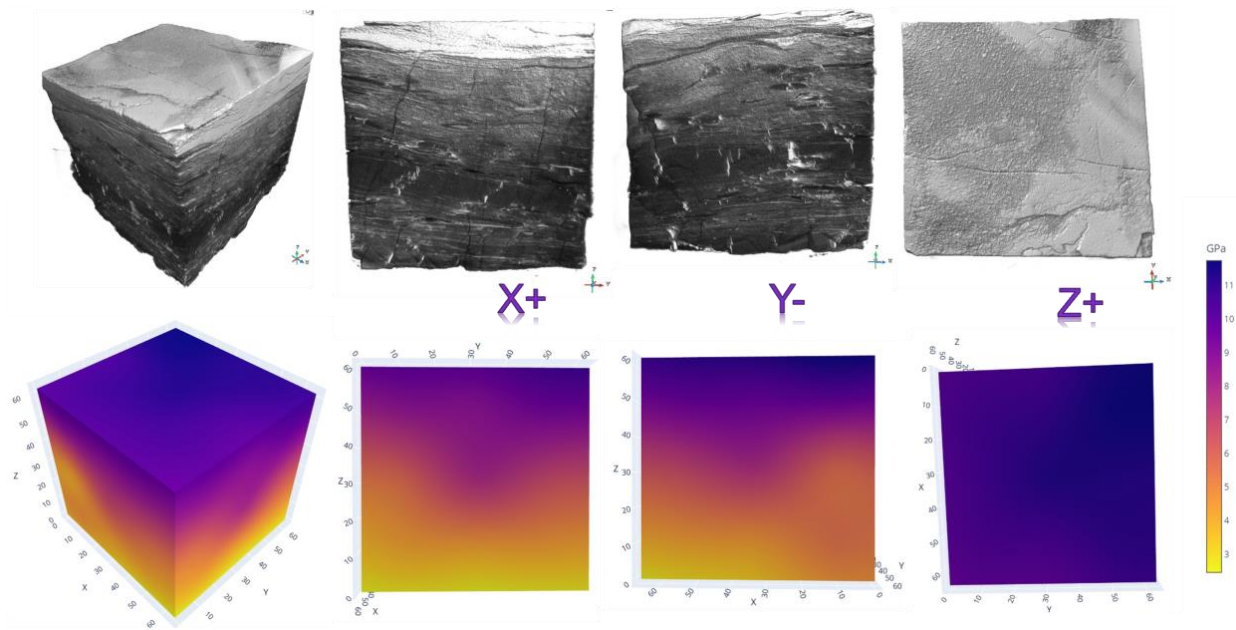

*Figure 8* *Comparison of the reconstructed digital coal sample and dynamic modulus mapping in both 3D and 2D representations for each direction (Example 2).*

## 4 Conclusions

The study introduces a novel, non-destructive sonic mapping technique for evaluating the internal mechanical properties and structural heterogeneity of coal samples. This method measures P-wave (Vp) and S-wave (Vs) velocities across three orthogonal axes (X, Y, Z) of cubic coal samples. These velocity data are used to generate detailed 2D maps for each axis and interpolate them to create a high-resolution 3D sonic map.

The method's accuracy is validated against high-resolution CT scan images, demonstrating its effectiveness in identifying fractures, voids, and material composition variations within coal. By correlating sonic velocity variations with density distribution observed in CT scans, the study confirms the reliability of the 3D sonic mapping approach.

This technique provides a simple, powerful, and non-invasive tool for material characterization and could have applications in fields such as geotechnical engineering, mining, construction, and material science.

The proposed sonic mapping method offers a simple yet robust framework for assessing the internal structure and mechanical properties of coal samples. Its non-destructive nature and validated accuracy through CT scans make it a promising technique for detecting structural heterogeneities like fractures and voids. By delivering detailed 3D insights, this approach enhances the precision of material evaluations, with potential applicability across various industries, including geology, construction, and advanced material testing.

## Data availability

The data used in this study and the python library "SonicMapping" developed in the study have been fully open sourced on:

https://github.com/lixuekai2001/SonicMapping